\documentclass[twocolumn,superscriptaddress,amsmath,aps,prl]{revtex4-2}
\usepackage[utf8]{inputenc}
\usepackage[final]{graphicx}
\DeclareGraphicsExtensions{.eps}
\usepackage{wrapfig}
\usepackage{subfigure}
\usepackage{bm}
\usepackage[normalem]{ulem}
\usepackage[usenames]{color}
\usepackage{times}
\usepackage{verbatim}
\usepackage{xcolor}
\usepackage{hyperref}

\begin{document}
\title{Critical Dynamics: multiplicative noise fixed point in two dimensional systems}
\author{Nathan O.  Silvano}
\email{nathanosilvano@gmail.com }
\affiliation{Departamento de F{\'\i}sica Te\'orica,
Universidade do Estado do Rio de Janeiro, Rua S\~ao Francisco Xavier 524, 20550-013  
Rio de Janeiro, RJ,  Brazil}
\author{Daniel G.  Barci}
\email{daniel.barci@gmail.com}
\affiliation{Departamento de F{\'\i}sica Te\'orica,
Universidade do Estado do Rio de Janeiro, Rua S\~ao Francisco Xavier 524, 20550-013  
Rio de Janeiro, RJ, Brazil}
\date{\today}

\begin{abstract}
We study the critical dynamics of a real scalar field in two dimensions near a continuous phase transition.   We have  built up and solved  Dynamical  Renormalization Group equations at one-loop approximation.  We have found that,  different form  the case $d\lesssim 4$,  characterized by a Wilson-Fisher fixed point with dynamical critical exponent $z=2+ O(\epsilon^2)$,  the critical dynamics is dominated by a novel multiplicative noise fixed point.  The zeroes of the beta function depend on the stochastic prescription used to define the Wiener integrals.  However,   the critical exponents and the anomalous dimension do not depend on the prescription used.  Thus,  even though each stochastic prescription produces different dynamical evolutions,   all of them are in the same universality class.   
\end{abstract}

\maketitle

Out-of-equilibrium evolution near continuous phase transitions is a fascinating subject. While equilibrium properties are strongly constrained by symmetry and dimensionality, the dynamics is much more involved  and generally depends on conserved quantities and other details of the system.  The interest in critical dynamics is rapidly growing up in part due to the  wide range of  multidisciplinary applications in which criticallity  has deeply impacted.  For instance, the collective behavior of  different biological systems has critical properties,  displaying  space-time correlation functions with non-trivial scaling laws\cite{TGrigera-2019,Gambassi2021}.  Other interesting examples come from epidemic spreading models where  dynamic percolation  is observed near multicritical  points\cite{Janssen-2004}.  Moreover,   strongly correlated systems,  such as antiferromagnets in transition-metal oxides~\cite{Cabra-2005,Bergman-2006} present a very rich phase diagram including ordered as well as topological phases. These compounds are generally described by  dimmer models or related quantum field theory models\cite{Hsu-2013},  that  seem to have  anomalous critical dynamics\cite{Isakov-2011}. 

From a theoretical point of view,  the standard approach to critical dynamics is the ``Dynamical Renormalization Group (DRG)"\cite{Bausch-1976},  distinctly developed  in  a seminal paper by Hohenberg and Halperin\cite{Hohenberg-Halperin-1977}.  The simplest starting point is to assume that,  very near a critical point,  the dynamics of the order parameter is governed by a dissipative processes driven by  an overdamped additive noise Langevin equation.  Then, the critical point is approached by  integrating out short distance (high momentum) degrees of freedom in order to obtain the dynamics of an equivalent effective long distance (small momentum) model.  
From this procedure,  it is possible to read,  for instance,  the typical relaxation time near a fixed point,  given by  $\tau\sim \xi^z$,  where $\xi$ is the correlation length and $z$ is the dynamical critical exponent.  At a critical point, $\xi\to\infty$ and therefore $\tau\to \infty$, meaning that the system does not reach the equilibrium at criticallity.  Together with usual equilibrium exponents,  $z$ defines the universality class of the transition.  Interestingly,  since the symmetry of the model does not constrain dynamics,  there are different dynamic universality classes for the same critical point.

As usual in renormalization group  theory (RG)\cite{Cardy-book},  the integration over higher momentum modes  generates all kind of couplings, compatible with the symmetry of the problem.  For this reason a consistent study of a RG flux should begin, at least formally, with the most general Hamiltonian containing all couplings compatible with symmetry.  Interestingly, in a similar way,  DRG transformations generate couplings that modify the probability distribution of the original stochastic process.  In particular, we will show that one-loop perturbative corrections generate couplings compatible with a multiplicative noise stochastic processes~\cite{vanKampen},  even in the case of assuming an additive processes as a starting point.  In order to understand the fate of these couplings,  we decided to  analyze a more general  dynamics for the order parameter near criticallity,  assuming a dissipative process driven by a  multiplicative noise Langevin equation.  

\begin{figure*}[htb]
\begin{center}
\subfigure[]
{\label{fig:FPur}
\includegraphics[width=0.25\textwidth]{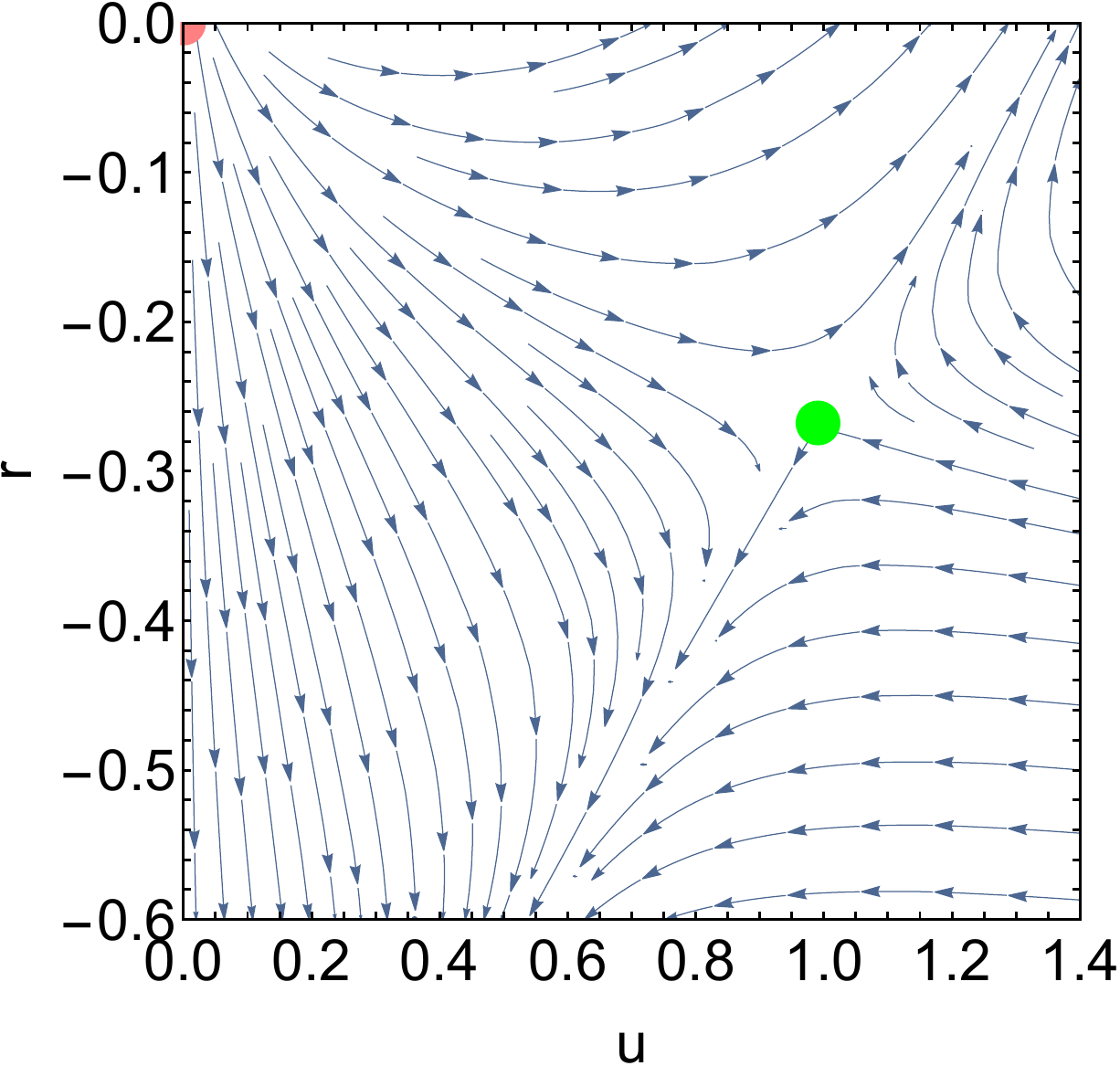}} 
\subfigure[]
{\label{fig:FPug}
\includegraphics[width=0.24\textwidth]{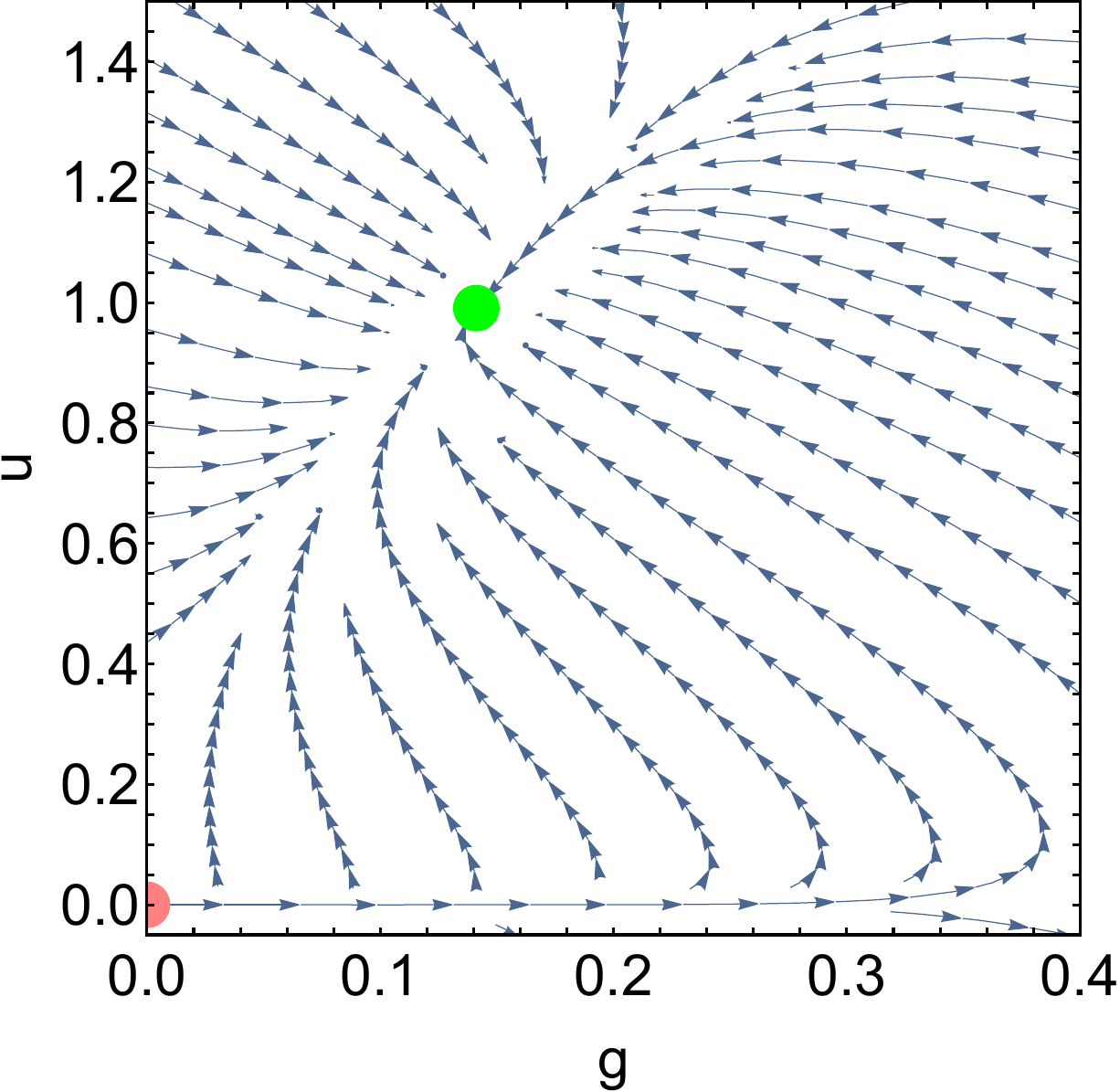}} 
\subfigure[]
{\label{fig:FPrg}
\includegraphics[width=0.25\textwidth]{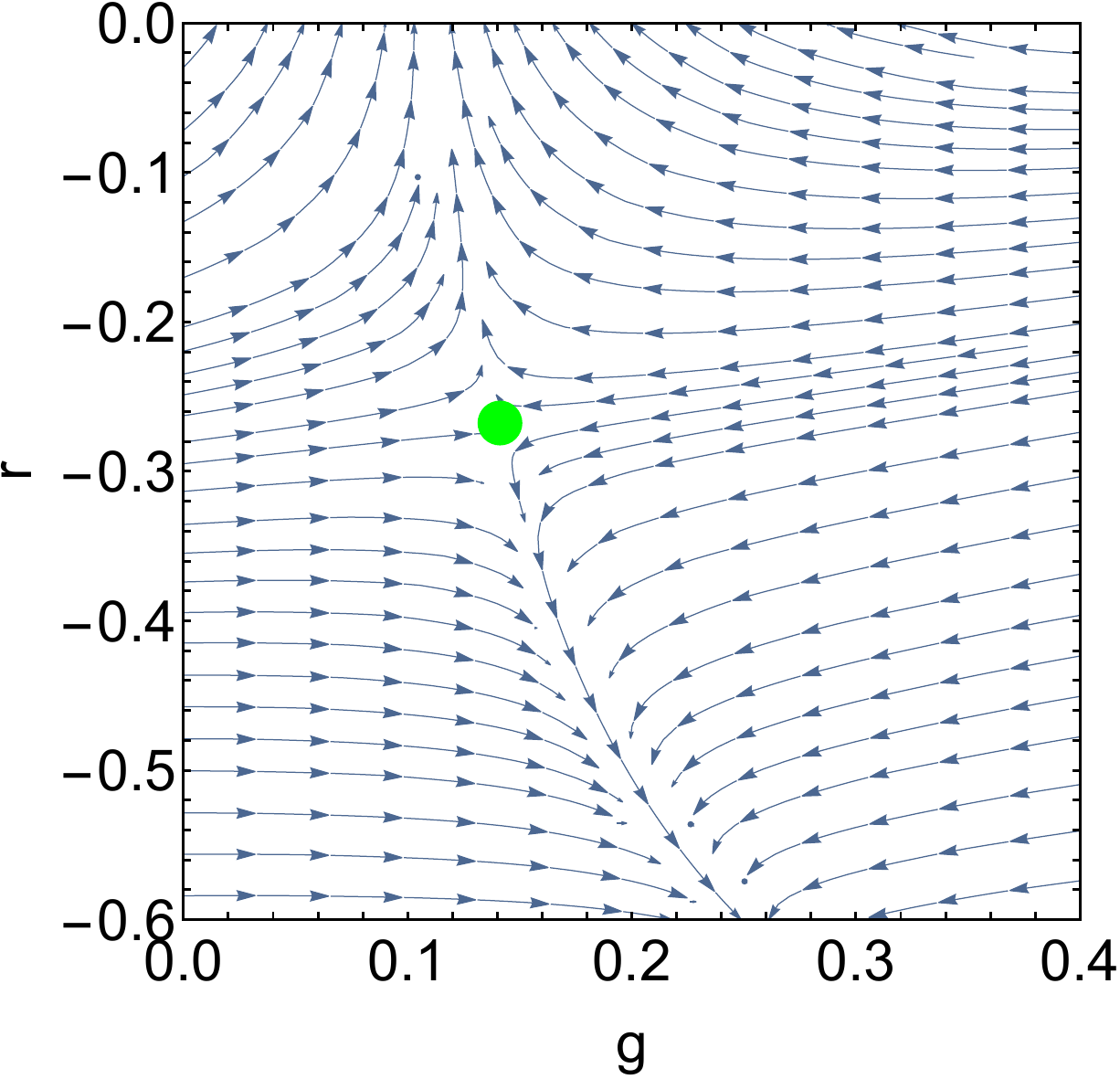}}
\end{center}
\caption{ DRG flux, obtained by numerically solving Equations \ref{eq:DRG-r} to \ref{eq:DRG-g}.  We have fixed the Stratonovich prescription $\alpha=1/2$.  In (a),  we depict the flux in the $r-u$ plane at $g^*=0.141$.  It can be clearly seen a fix point $u^*=0.991$,  $r^*=-0.268$,  with  one attractive and one repulsive direction with a similar topology of a Wilson-Fischer fixed point.   In (b) we show the $u-g$ plane at $r^*=-0.268$.  We can observe  a completely attractive fixed point at $u^*=0.991$, $g^*=0.141$.    In (c),  we show the fixed point  in the $r-g$ plane $r^*=-0.268$,$g^*=0.1411$  at the critical value $u^*=0.991$.}
\label{fig:DRGfluxes}
\end{figure*}

For concreteness,  we present a simple model of a not conserved real scalar order parameter, $\phi({\bf x},t)$ with quartic coupling $\phi^4({\bf x},t)$ (model A of Ref. \onlinecite{Hohenberg-Halperin-1977}).  We assume  a  multiplicative noise Langevin equation,  modeled by a  general  dissipation function $G(\phi)$,  with the same symmetry of the Hamiltonian.
The upper critical dimension of this model  is $d_c=4$.   For $d>4$,  the Gaussian fixed point with $z=2$ correctly describes the phase transition.  For $d\lesssim  4$,  the Gaussian fixed point turns out to be  unstable and the Wilson-Fisher fixed point\cite{Wilson-1972}  shows up in a first order expansion around $\epsilon=4-d$.  In this case,  the dynamics is governed by  $z=2+O(\epsilon^2)$ and all multiplicative noise coupling constants are irrelevant.  In this sense,  we recover the very well known results of Ref.  \onlinecite{Hohenberg-Halperin-1977}.  However, the dynamical behavior dramatically changes at $d=2$.    The main result of this letter is that,  at $d=2$,  all multiplicative noise couplings are {\em marginally relevant}, 
flowing to a novel stable fixed point dominated by a multiplicative stochastic process.    
In figure  \ref{fig:DRGfluxes},  we show the novel fixed point as well as the DRG flux.   
In the following,  we present the model,  we give details of  the  calculations and finally we discuss our results.

{\em Model--} We consider the following Hamiltonian of a real scalar field $\phi({\bf x})$: 
\begin{equation}
H=\int_{\frac{1}{\Lambda}} d^d x \left\{\frac{1}{2}\left| {\bf \nabla}\phi\right|^2+\frac{r}{2}\phi^2+\frac{u}{4!}\phi^4\right\}
\label{eq:Hamiltonian}
\end{equation}
where $\Lambda$ is an ultraviolet momentum cut-off. $\{r,u\}$ are the quadratic and quartic coupling constants respectively. 
The dynamic evolution is given by the Langevin equation, 
\begin{equation}
\frac{\partial\phi({\bf x},t)}{\partial t}=-\Gamma \frac{\delta H}{\delta \phi({\bf x},t)}+ G(\phi^2)\eta({\bf x},t)
\label{eq:Langevin}
\end{equation}
where $\eta({\bf x}, t)$ is a Gaussian white noise field:
$\langle \eta({\bf x},t) \rangle=0$ and  
$\langle \eta({\bf x},t)\eta({\bf x}',t') \rangle=2\Gamma\delta({\bf x}-{\bf x}')\delta(t-t')$.
$\Gamma$ is a diffusion constant  and $G(\phi^2)$ is a diffusion function characterizing the multiplicative noise distribution.  $\langle\ldots\rangle$ represents stochastic mean value.    In order to correctly define  Eq. (\ref{eq:Langevin}), we need to fix a stochastic prescription to properly compute the time Wiener integrals.  In this paper,  we adopted the so called Generalized Stratonovich prescription  that is parametrized by a real number $0\leq\alpha\leq 1$,  in such a way that different values of $\alpha$ correspond to specific prescriptions.  For instance,  $\alpha=0$ is the It\^o prescription, $\alpha=1/2$ is the Stratonovich one, 
while $\alpha=1$ is the Hangii-Klimontovich or anti-It\^o convention\cite{Arenas2012}.  

To compute dynamical correlation functions,  we use a generalized  Martin-Siggia-Rose-Janssen-DeDominicis  formalism  (MSRJD)~\cite{MSR1973,Janssen-1976,deDominicis-1976},  that represents the Langevin dynamics by a field theory.  The generating functional~\cite{Arenas2012,Arenas2012-2, Miguel2015}
\begin{equation}
Z[J_\phi,J_\varphi]=\int {\cal D}\phi{\cal D}\varphi {\cal D}\bar\xi{\cal D}\xi\; e^{-S[\phi,\varphi,\bar\xi,\xi]+\int d^dxdt \{J_{\phi}\phi+J_{\varphi}\varphi\}}
\label{eq:ZJ}
\end{equation}
is written as a functional integral over four fields, where $\phi({\bf x}, t)$ and $\varphi({\bf x}, t)$ are two real scalar fields essentially representing a local order parameter and a response field respectively.  $\bar\xi({\bf x}, t)$ and $\xi({\bf x}, t)$ are anticommuting Grassmann fields. 

The ``action'' is given by, 
\begin{align}
&S=\int_{\frac{1}{\Lambda}} d^dx dt\;  i\varphi \left\{\frac{\partial\phi}{\partial t}+\frac{\Gamma}{2} \frac{\delta H}{\delta \phi}
+\Gamma G\frac{\delta G}{\delta\phi}\bar\xi \xi\right\}+\frac{\Gamma}{2} G^2\varphi^2
\nonumber \\
&-\int_{\frac{1}{\Lambda}} d^dx dtd^dx' dt'\;  \bar\xi({\bf x},t)K({\bf x}-{\bf x}', t-t')\xi({\bf x},t) 
\label{eq:S}
\end{align}
where the kernel 
\begin{equation}
K({\bf x}-{\bf x}', t-t')=
\frac{d\delta(t-t')}{dt}+\frac{\Gamma}{2}\frac{\delta^2 H}{\delta\phi({\bf x},t)\delta\phi({\bf x}',t)}
\label{eq:K}
\end{equation}

Without loose generality, we chose a diffusion function satisfying $G(0)=1$, which is the usual additive noise value. Then, we can write
\begin{equation}
G^2(\phi^2)=1+\sum_{n=1}^{\infty} g_n \phi^{2n} 
\label{eq:G2}
\end{equation}
where $g_n$, with $n=1,2,\ldots$, are coupling constants defining the multiplicative noise distribution.

By trivial power counting, we immediately verify that $[r]=2$, $[u]=4-d$,   $[\Gamma]=z-2$ and $[g_n]=n(2-d)$,where the notation $[\ldots]$ indicates dimension in powers of momentum.  Of course, the dimensions of $r$ and $u$ are the usual ones  in the equilibrium theory.  This defines the upper-critical dimension $d_c=4$.   The dynamical critical exponent $z$ is fixed by demanding $[\Gamma]=0$.    For  $d>2$,  $[g_n] <0$,  therefore,  all the multiplicative noise couplings are irrelevant at tree level.  However,  at $d=2$,  $[g_n]=0$, $\forall n$. In this case, the entire function $G[\phi]$ produces marginal couplings at tree level.   In order to study the fate of the multiplicative coupling,  it is necessary to compute fluctuations. 

Let us study in detail the simpler model of just one marginal coupling constant $g_1=g$, and $g_n=0$ for $n\neq 1$.  In this case, $G^2=1+g \phi^2$.  We can split the action into two terms,   $S=S_0+S_I$,  where $S_0$ is the quadratic part and $S_I$ codifies  interaction terms.  We find,  
\begin{equation}
S_0=\int_{\frac{1}{\Lambda}} d^dx dt\; \left\{ i\varphi \Delta \phi + \frac{\Gamma}{2} \varphi^2-\bar\xi \Delta \xi\right\}
\label{eq:S0}
\end{equation}
where the differential operator $\Delta=\partial_t+(\Gamma/2) (-\nabla^2+r)$.
The interacting part of the action reads, 
\begin{align}
S_I&=\int_{\frac{1}{\Lambda}} d^dx dt\;  \times
\label{eq:SI} \\
&\times\left\{ \frac{iu\Gamma}{3! 2}\varphi \phi^3-\frac{u\Gamma}{4} \phi^2\bar\xi\xi
+\frac{g\Gamma}{2} \varphi^2\phi^2+ig\Gamma\varphi\phi\bar\xi\xi\right\} \; .
\nonumber
\end{align}
The last two terms codify the effect of multiplicative noise dynamics.

{\em DRG transformation--} In order to built a DRG transformation we firstly lower the ultraviolet cut-off to $\Lambda/b$, with $b>1$ and split the fields: $\phi=\phi^<+\phi^>$, $\varphi=\varphi^<+\varphi^>$, $\bar\xi=\bar\xi^<+\bar\xi^>$ and $\xi=\xi^<+\xi^>$,  in such a way that the Fourier transformed  fields with superscript ``$<$",  have support within the sphere $k<\Lambda/b$, while the fields labeled by ``$>$" have support on a spherical shell $\Lambda/b<k<\Lambda$.  We define the transformed action $S'_{\Lambda/b}[\phi^<,\varphi^<,\bar\xi^<,\xi^<]$, by integrating out the fields with momentum higher than $\Lambda/b$ in the following way
\begin{equation}
e^{-S'_{\Lambda/b}}=\int {\cal D}\phi^>{\cal D}\varphi^> {\cal D}\bar\xi^>{\cal D}\xi^>\; e^{-S_\Lambda[\phi,\varphi,\bar\xi,\xi]}
\label{eq:DRGT}
\end{equation}
Since the transformed action, $S'_{\Lambda/b}$,  has a momentum cut-off $\Lambda/b$,  in order to compare it with the original one, $S_{\Lambda}$,  with cut-off $\Lambda$,  we re-scale momentum, frequency and fields as 
\begin{align}
k'=b k &\mbox{~,~} \omega'= b^z \omega \\
\varphi^<(k'/b, \omega'/b^z)&= b^{\frac{d-2+2z}{2}-\eta_1}\varphi(k', \omega') \\
\phi^<(k'/b, \omega'/b^z)&= b^{\frac{d+2+2z}{2}+\eta_2}\phi(k', \omega') \\
\bar\xi^<(k'/b, \omega'/b^z)&= b^{\frac{d}{2}+\eta_3}\bar\xi(k', \omega') \\
\xi^<(k'/b, \omega'/b^z)&= b^{\frac{d}{2}+\eta_3}\xi(k', \omega')
\end{align}
where $z$ is the dynamical critical exponent and $\eta_1,\eta_2,\eta_3$ are anomalous dimensions. 
 By comparing the coupling constants  in $S'$ with $S$ and, considering an infinitesimal transformation $b=1+\delta\ell$, with $\delta\ell<<1$, we find DRG differential equations for $\{r(\ell),u(\ell),\Gamma(\ell),g(\ell)\}$.

\begin{figure}
	\begin{center}
		\includegraphics[width=0.35\textwidth]{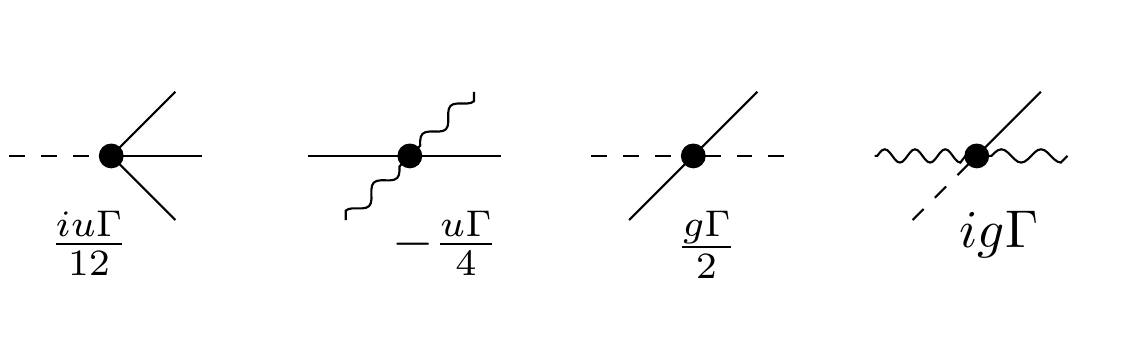}
	\end{center}
	\caption{Vertex as read from the  action $S_I$ of Eq. (\ref{eq:SI}).  The dashed line represent the field $\varphi({\bf x}, t)$ while the solid line $\phi({\bf x}, t)$.  The wiggle lines corresponds to the Grassmann fields $\{\bar\xi({\bf x}, t),
	\xi({\bf x}, t)\}$.  The last two vertex characterize the multiplicative noise distribution. }
	\label{fig:vertex}
\end{figure}

{\em Perturbative one-loop DRG--}
The main difficulty of  this procedure is to compute the functional integral of Eq. (\ref{eq:DRGT}). To do this, we implement a perturbative calculation. To second order in the couplings  we find,  
\begin{equation}
S'_{\Lambda/b}=  S^<+ \langle S_I\rangle_{S_0^>}-\frac{1}{2} \left(\langle S_I^2\rangle_{S_0^>}-\langle S_I\rangle_{S_0^>}^2\right)
\end{equation}
where $S^<=S_0^<+S_I^<$ means the action of Eqs. (\ref{eq:S0}) and (\ref{eq:SI}) with the substitution of the fields $\phi,\varphi,\bar\xi,\xi$ by $\phi^<,\varphi^<,\bar\xi^<,\xi^<$.  The notation $\langle\ldots\rangle_{S_0^>}$ means the mean value computed with the quadratic action $S_0$ of Eq. (\ref{eq:S0}) but with the momentum constrained to the small shell  $\Lambda/b<k<\Lambda$.  The more efficient way to organize this expansion is by using Feynman diagrams and grouping terms by the number of internal loops.
In fig.  \ref{fig:vertex} we depicted  the four vertex of our model. 
From $S_0$ we can read two propagators: $G^>({\bf x}-{\bf x}',t-t')=\langle \phi({\bf x},t)\phi({\bf x}',t')\rangle$,  represented by an internal continuous line
and $R^>({\bf x}-{\bf x}',t-t')=\langle \phi({\bf x},t)\varphi({\bf x}',t')\rangle\Theta(t-t')$, represented by an internal continuous line with an arrow.  The Grassmann field propagator is related with the response function $\langle \bar\xi({\bf x},t)\xi({\bf x}',t')\rangle=i R^>({\bf x}-{\bf x}',t-t')$ and is represented by an internal wiggled  line.  The bare response fields are not correlated $\langle \varphi({\bf x},t)\varphi({\bf x}',t')\rangle=0$.  In Fourier space,  the propagators read,   $\tilde G^>({\bf k},\omega)=
\Gamma[\omega^2+\frac{ \Gamma^2}{4} \left(k^2+r\right)^2]^{-1}$ and $R^>({\bf k},\omega)=
-i[-i\omega+ \frac{\Gamma}{2} \left(k^2+r\right)]^{-1}$.   In these expressions $1-\delta\ell<k/\Lambda <1$.

\begin{figure}
	\begin{center}
\includegraphics[width=0.35\textwidth]{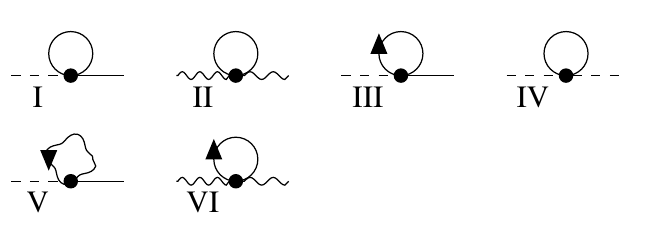}
	\end{center}
	\caption{First order corrections.  Internal continuous lines represent the $G^>$ propagator, while internal arrowed lines correspond to  the response function $R^>$.   Wiggle internal lines represent Grassmann propagators.  The diagrams I, II, III, V and VI are contributions that essentially renormalize the $r$ parameter.  Diagram IV,  contribute to renormalize $\Gamma$, been responsible for the anomalous dimension $\eta$. }
	\label{fig:1V1loop}
\end{figure}

\begin{figure}
	\begin{center}
\includegraphics[width=0.35\textwidth]{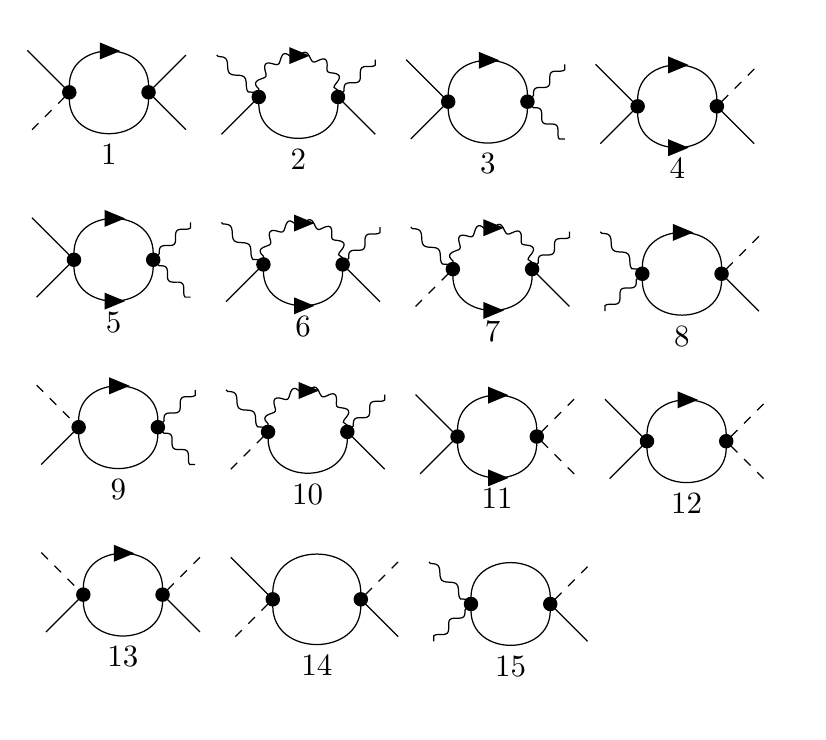}
	\end{center}
	\caption{One loop second order diagrams.  The line conventions are the same of fig. \ref{fig:1V1loop} }
	\label{fig:2V1loop}
\end{figure}

The first order corrections are depicted in  fig.  \ref{fig:1V1loop}, 
while we show the relevant second order one loop diagrams in figure \ref{fig:2V1loop}.
We explicitly computed all  diagrams and,  after rescaling momenta, frequencies and fields,  we can read the running coupling constants by comparing $S'_{\Lambda}$ with the original action $S_{\Lambda}$. 

The one-loop perturbative DRG equations are  displayed  in a simpler way  by using dimensionless coupling constants.   To this purpose we  rescaled the couplings $r\to  r\Lambda^{-2}$,   $u\to   (\Omega_d/(2\pi)^d) u\Lambda^{d-4}$ and $g\to (\Omega_d/(2\pi)^d) g \Lambda^{2-d}$,   where $\Omega_d$  is the area of a $(d-1)$ dimensional sphere or radius one.  We have found:
\begin{align}
\frac{dr}{d\ell}&= 2r+\frac{1}{2}\frac{u}{1+r}- 2\alpha g 
\label{eq:DRG-r}\\
\frac{du}{d\ell}&= (4-d) u-\frac{3}{2}\frac{u^2}{(1+r)^2}+4 \frac{u g}{1+r}
\label{eq:DRG-u} \\
\frac{d\Gamma}{d\ell}&=\left\{z-2 \right\}  \Gamma
\label{eq:DRG-Gamma}   \\
\frac{dg}{d\ell}&= (2-d) g+\frac{g^2}{1+r}-\frac{5}{2}\frac{ g u}{(1+r)^2} +\frac{1}{4}\frac{u^2}{(1+r)^3}
\label{eq:DRG-g}
\end{align}
with the anomalous dimensions,  $\eta\equiv\eta_1=\eta_2=g/[2(1+r)]$ and $\eta_3=0$.
The first three equations, with $g=0$,  are the usual DRG equation at one-loop approximation.  Multiplicative noise contributes  with  third term  of Equation (\ref{eq:DRG-r}),  coming from diagrams III and V  of figure \ref{fig:1V1loop}.  In these diagrams,   the self-correlated  response function is  singular, therefore,  it is necessary to define it as $R^>_{\rm retarded}(0,0)= -i \alpha$,   where the parameter $\alpha$ corresponds with the stochastic prescription\cite{Arenas2012-2},  used to defined the Langevin equation (\ref{eq:Langevin}).  Multiplicative noise also contribute to the evolution of $u$,   as can be seen in Eq. (\ref{eq:DRG-u}).  The last term,  proportional to $ug$,  is generated by diagram 4 of figure \ref{fig:2V1loop}. 
 Eq. (\ref{eq:DRG-g}) is the key equation that controls the fate of the multiplicative coupling.  The last three terms come from diagrams 11, 12,13 and 14 of figure \ref{fig:2V1loop}.

This system has several fixed points, depending essentially on dimensionality. For $d>4$,  the Gaussian fixed point, $u^*=r^*=g^*=0$  with $z=2$ correctly describes the phase transition.  For $d\lesssim 4$, a Wilson-Fisher fixed point shows up at order $\epsilon=4-d$,  $u^*=(2/3) \epsilon$, $r^*=-(1/6) \epsilon$ with $z=2+O(\epsilon^2)$ and $g^*\sim O(\epsilon^2)$.  At this level of approximation, $g$ is an irrelevant scaling variable and the dynamics is driven by a usual additive noise stochastic process.  We have checked that this behavior remains the same at $O(\epsilon^2)$  thus,  we recover the well known results of Ref. \cite{Hohenberg-Halperin-1977}.  However, at $d=2$ the dynamical behavior completely changes its character. 
In this case, $g$ is no longer irrelevant but turns out to be {\em marginally relevant}. The flux gets away from its additive value,  even in the case of having an initial value $g=0$.  We have  found a novel fixed point 
$r^*=[\alpha  \left(-4 \alpha -3 \sqrt{21}+17\right)+2 \left(\sqrt{21}+3\right)]/(4 (\alpha -6) \alpha -48)$,   $u^*=[\alpha  \left( [49+11 \sqrt{21}] \alpha +32 \sqrt{21}+168\right)+36 \sqrt{21}+84]/[{2 ((\alpha -6) \alpha -12)^2}]$,  $g^*=[-\alpha +\sqrt{21}+3]^{-1}$.  The position of this fixed point is weakly dependent on the stochastic prescription $\alpha$.  In the Stratonovich prescription,  $\alpha=1/2$, the fixed point is located at $r^*=-0.268$,  $u^*=0.991$  and $g^*=0.141$.   Linearizing the DRG equations around this fixed point,  we have one repulsive and two attractive directions defined by the eigenvalues  $y=\{2, -6.002,-1.926\}$, with the corresponding normalized  eigenvectors $v_1=\{0.594,  0.804, 0 \}$,  $v_2=\{0.142, -0.915,  0.377 \}$ e $v_3=\{-0.119, 0.949, 0.290\}$.  These results are more easily visualized in figure \ref{fig:DRGfluxes},  where we show the DRG flux obtained by  numerically solving equations (\ref{eq:DRG-r})-(\ref{eq:DRG-g}).   The relevant eigenvalue $y=2$ as well as the anomalous dimension $\eta=2/(3 \sqrt{21}+7)$  do not depend on $\alpha$.   

{\em Summary and discussions.}
We have built up and solved the DRG equations of a real scalar field with quartic interactions whose dynamics is driven by a general overdamped Langevin equation.  The DRG transformation generates couplings that modify the distribution probability of the stochastic process.  In particular, multiplicative noise processes are generated by integrating out higher momentum degrees of freedom.  The main result of the letter is that,  at  $d=2$,   multiplicative couplings are {\em marginally relevant},  driven the DRG flux to a novel fixed point dominated by a not trivial  multiplicative noise process.      
The zeroes of the $\beta-$function explicitly depend on the stochastic prescription.  However, and most importantly,   the critical exponent $\nu=1/2$ and the anomalous dimension $\eta\sim 0.096$  do not depend on the value of $\alpha$.   This means that dynamical evolutions driven by different stochastic prescriptions  are in the same universality class.  It is worth noting that,  at one-loop level,  the anomalous dimension enters the fields $\phi$ and $\varphi$ with different sign.  This fact produces  non trivial power law decay of $\langle\phi\phi\rangle$ correlations,  however it does not modify the asymptotic behavior of  the response function  $\langle\phi\varphi\rangle$.  This fact is modified at two loop level,  in which $\eta_1\neq\eta_2$.    
 
At this point,  some caveats  are in order.   Perturbation theory is not a quite controlled approximation in two-dimensions.  Even though we expect  substantial corrections in  the position of the fixed points and the critical exponents,  we strongly believe that the existence of the multiplicative noise fixed point and the topology of the   
DRG flux is robust.  We have computed two-loop corrections to the DRG equations  and have found similar qualitative results.  We will present  details of this calculation elsewhere.  Moreover,  several exhaustive calculations of the equilibrium RG in $d=2$,  up to five loops approximation\cite{Baker-1978,Orlov-2000,Sokolov-2006}  further support this claim.  In order to get a deeper understanding on the fate of the multiplicative  dynamical fixed point  at higher orders and to precisely compute critical exponents,  it could be interesting to implement a non-perturbative approach to the DRG\cite{Dupuis2021}. 

\begin{acknowledgments}
The Brazilian agencies, {\em Funda\c c\~ao de Amparo \`a Pesquisa do Estado do Rio
de Janeiro} (FAPERJ), {\em Conselho Nacional de Desenvolvimento Cient\'\i
fico e Tecnol\'ogico} (CNPq) and {\em Coordena\c c\~ao  de Aperfei\c coamento de Pessoal de N\'\i vel Superior}  (CAPES) - Finance Code 001,  are acknowledged  for partial financial support.  NS was partially supported by a  PhD Fellowship from CAPES. 
\end{acknowledgments}

%

\end{document}